# Free-standing bialkali photocathodes using atomically thin substrates

*Hisato Yamaguchi[#,*], Fangze Liu[#], Jeffrey DeFazio, Mengjia Gaowei, Claudia W. Narvaez Villarrubia, Junqi Xie, John Sinsheimer, Derek Strom, Vitaly Pavlenko, Kevin L. Jensen, John Smedley, Aditya D. Mohite, Nathan A. Moody[*]*

Dr. H.Y., Dr. F.L., Dr. C.W.N.V., Dr. A.D.M.
MPA-11 Materials Synthesis and Integrated Devices (MSID), Materials Physics and Applications Division, Mail Stop: K763, Los Alamos National Laboratory, P.O. Box 1663, Los Alamos, New Mexico 87545, U.S.A.
E-mail: hyamaguchi@lanl.gov
Dr. J.D.
Photonis USA Pennsylvania Inc., 1000 New Holland Ave., Lancaster, PA 17601, U.S.A.
Dr. M.G., Dr. J.S., Dr. J.Smedley
Brookhaven National Laboratory, P.O. Box 5000, Upton, New York 11973, U.S.A.
Dr. J.X.
Argonne National Laboratory, 9700 South Cass Ave., Argonne, Illinois 60439, U.S.A.
Dr. D.S.
Max Planck Institute for Physics, Föhringer Ring 6, D-80805 Munich, Germany
Dr. V.P., Dr. N.A.M.
Accelerators and Electrodynamics (AOT-AE), Accelerator Operations and Technology Division, Mail Stop: H851, Los Alamos National Laboratory, P.O. Box 1663, Los Alamos, New Mexico 87545, U.S.A.
E-mail: nmoody@lanl.gov
Dr. K.L.J.
Materials and Systems Branch (Code 6360), Materials Science and Technology Division, Naval Research Laboratory, Washington DC 20375, U.S.A.

[#]These authors contributed equally.

**Abstract**

We report successful deposition of high quantum efficiency (QE) bialkali antimonide $K_2CsSb$ photocathodes on graphene films. The results pave a pathway towards an ultimate goal of encapsulating technologically-relevant photocathodes for accelerator technology with an atomically thin protecting layer to enhance lifetime while minimizing QE losses. A QE of 17 % at ~3.1 eV (405 nm) is the highest value reported so far on graphene substrates and is comparable to that obtained on stainless steel and nickel reference substrates. The spectral responses of the photocathodes on graphene exhibit signature features of $K_2CsSb$ including the characteristic absorption at ~2.5 eV. Materials characterization based on X-ray fluorescence (XRF) and X-ray diffraction (XRD) reveals that the composition and crystal quality of these photocathodes deposited on graphene is comparable to those deposited on a reference substrate. Quantitative agreement between optical calculations and QE measurements for the $K_2CsSb$ on free suspended graphene and a graphene coated metal substrate further confirms the high quality interface between the photocathodes and graphene. Finally, a correlation between the QE and graphene quality as characterized by Raman spectroscopy suggests that a lower density of atomistic defects in the graphene films leads to higher QE of the deposited $K_2CsSb$ photocathodes.

# 1. Introduction

Many grand challenges for humanity, including the quest for sustainable energy, continued scaling of computational power, detection and mitigation of pathogens, and study of the structure and dynamics of the building blocks of life require the ability to access, observe, and control matter on the frontier timescale of electronic motion and the spatial scale of atomic bonds[1, 2]. The only instruments with such capabilities are future coherent x-ray sources and advanced colliders, which demand increasingly high performance electron beams[3-6]. The performance requirements dramatically outstrip the capabilities of present state-of-the-art electron sources and cathode technologies[7]. The need for breakthrough advances in electron sources has recently come into sharp focus. Studies commissioned by the U.S. department of energy (DOE) have repeatedly identified electron sources as a critical risk area, forming one of the highest accelerator R&D priorities for the next decade, requiring transformational advance of cold cathode performance in particular[3, 4, 6, 8].

The required advances in cold cathode performance requires that radical improvements in lifetime and efficiency be achieved *simultaneously*. Previous work has succeeded in delivering increases in one, but at considerable expense of the other due to the competing physical processes underlying traditional approaches to cathode design and optimization[9-18]. In pursuit of this goal, novel properties that arise from unique geometry/dimensionality and size confinement effects of nanomaterials are expected to make an important contribution. The unique approach started here aims to decouple the competing mechanisms so that both high efficiency and long lifetime can be achieved via integration of atomically thin two dimensional (2D) nanomaterials with high-performing existing photocathode technologies. The generalized hypothesis driving this approach is that lifetime can be significantly enhanced without sacrificing the quantum efficiency (QE) by protecting photocathodes with atomically thin materials, yielding emergent properties for controlled functionality of cathode properties. A particular material combination of interest is the bialkali antimonide semiconductor $K_2CsSb$ and graphene. These photocathodes of choice possess one of the highest QEs with a peak well exceeding 20 % at 3 eV. The only better established photocathode known is activated gallium arsenide (GaAs: Cs-O), however this cathode material requires extremely high operating vacuum ~$10^{-11}$ Torr[7]. Therefore the $K_2CsSb$ forms a better choice for an investigation of compatibility with other systems such as 2D materials. Regarding a protecting layer, graphene is known to possess an exceptionally high gas barrier property despite being atomically thin[19-21]. This atomic thinness is crucial because it offers a unique opportunity for minimal sacrifice to the QE of encapsulated photocathodes[22]. Other advantageous material properties of graphene that are relevant to these demanding applications are: ultra-high electrical and thermal conductivity, optical transparency, high charge mobility, and ability to sustain extreme current densities[23-33].

Our prior reports on these efforts have shown the relevance and promise of graphene in photocathode design, emphasizing its potential as a unique substrate and passivating barrier[34, 35]. In this report, we made significant progress in the integration of few layer graphene with $K_2CsSb$. Specifically, state-of-the-art photocathodes (traditionally grown on thick substrates) can now be grown on vanishingly thin transparent substrates made of few layer graphene with QE that is comparable to those deposited on well-established rigid substrates. This configuration serves as a milestone toward ultimate goal of encapsulating high performance but environmentally susceptible photocathodes with graphene as passivating barrier.

## 2. Results and Discussion
### 2.1 Bialkali antimonide photocathodes on graphene substrates in a sealed vacuum tube
*2.1.1 Graphene substrate synthesis and photocathode deposition*

The graphene used in this study was grown by chemical vapor deposition (CVD) and was confirmed to be monolayer with minimal structural defects. Specifically, Raman spectroscopy showed a 2D/G peak ratio of ~3, where a 2D/G value of higher than ~2 is accepted as an indication of a monolayer[36]. There was no observable D peak at ~1350 cm$^{-1}$ (Supporting Information) that indicates the structural defect induced vibration mode in graphitic materials thus demonstrating there are only minimal defects in our graphene. Further characterization by atomic force microscopy (AFM), scanning electron microscopy (SEM), and optical microscopy was consistent with that of Raman spectroscopy. AFM showed continuous films with monolayer thickness of ~0.5 nm, and SEM and optical microscopy showed film uniformity over a range of ~7 mm (Supporting Information). These CVD graphene films were stacked and transferred onto metal mesh grids using an established polymer-supported wet-based method. A stacking of 3 monolayers of graphene prior to the transfer resulted in nearly 100 % coverage of the 34 μm diameter holes in the mesh grids. After the removal of the polymer-support in an acetone bath and drying, the graphene-spanned grids were installed into vacuum tube assemblies where bialkali antimonide photocathodes were deposited on the films and permanently sealed. The vacuum phototube setting allows a unique opportunity for long-term stability that can be inaccessible in dynamic pumping environments, and the design used here allows for routine and repeatable quantum efficiency measurements of photocathodes on the various graphene substrates (**Figure 1** (a)). Note that the graphene substrate is located between the metal mesh grid and photocathode film (Figure 1 (c)). This unique configuration allows us to investigate the QE of photocathodes deposited on suspended graphene (i.e. spanning the hole region). The results provide a critical evaluation on the photocathode quality that is interfaced intrinsically with an atomically thin protecting layer, which aligns well with our ultimate goals.

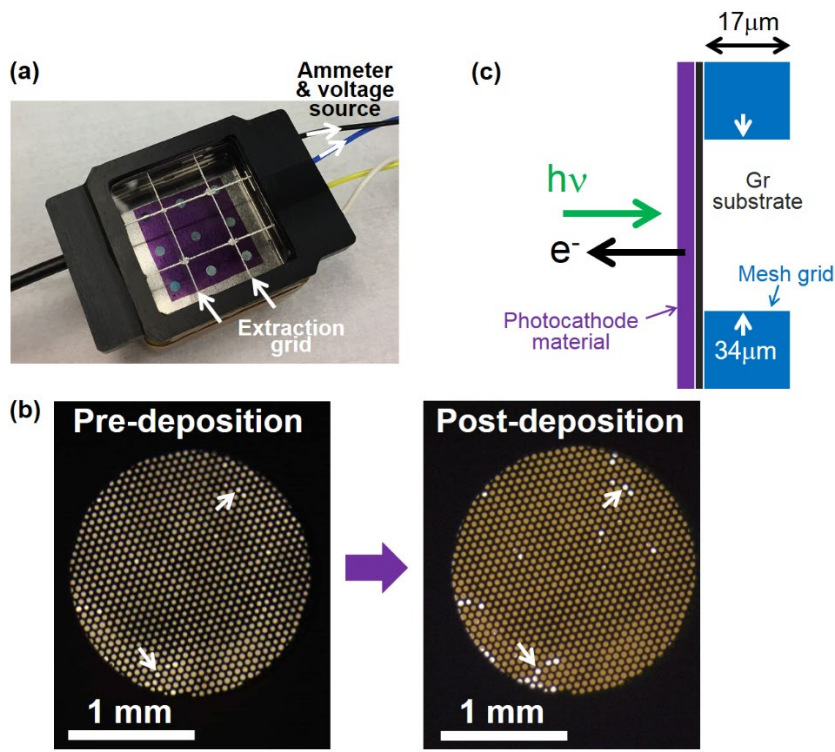

**Figure 1.** (a) Photograph of graphene substrate photocathodes that are deposited and sealed under vacuum. (b) Optical miscopy images of graphene substrates on SS mesh grids prior to (left) and after (right) $K_2CsSb$ photocathode deposition (graphene side). White arrows indicate void regions. Scale bars are 1 mm. (c) Schematic of graphene substrate photocathode structure and photoemission measurements performed in this study.

By carefully considering the mounting scheme any mechanical deformations of the extremely delicate graphene-spanning grids were minimized, and as a result a very low void density of ~ 1.5 % (Figure 1 (b)) was achieved. After the photocathode growth it is already apparent from the optical microscopy that photocathode material is indeed fully spanning the open areas without causing damage. Considering the typical $K_2CsSb$ photocathode film thickness of > 20 times that of the freely suspended graphene, this is already a promising indication for compatibility of these materials.

*2.1.2 Quantum efficiency maps*

**Figure 2** (a) and (b) are quantum efficiency (QE) maps obtained by rastering a 405 nm (~3.1 eV) light emitting diode (LED) with spot size of ~0.2 mm over the ~2 mm diameter sample areas. The samples are characterized in reflection mode, i.e. illuminated from photocathode sides with photoemission current collected from the same side (Figure 1(c)). A notably high QE approaching 17 % (16.6 % maximum, 16.1 % mean) over a large area ~ 2 mm diameter from the $K_2CsSb$ photocathode on graphene substrates was observed (Figure 2 (a)), and similar behavior of high QE and large area uniformity was confirmed for at least three additional samples. As a comparison, the $K_2CsSb$ photocathode was simultaneously deposited on pure Ni foil and chemically passivated stainless steel (SS). The QE on Ni had a mean value of 14.2 %, which is slightly lower than that of $K_2CsSb$ on the graphene samples. Moreover, the QE was notably less uniform for the Ni substrate as can be seen in Figure 2b and 2c. The SS, which is an established metal substrate for bialkali photocathode growth, exhibited QE (19.3 % mean) that are higher than that of graphene with distribution widths comparable to that of the graphene samples[7]. The QE in Figure 2 (a) however does not account for difference in QE between the suspended and supported regions of the graphene substrates. Differences in the net optical absorption of photocathodes between the supported (opaque/reflective) and the suspended (semitransparent) regions of graphene are effectively combined to an average in these measurements, whereas the suspended regions are of primary interest given our ultimate goal of enhancing the lifetime of photocathodes with atomically thin protecting layers.

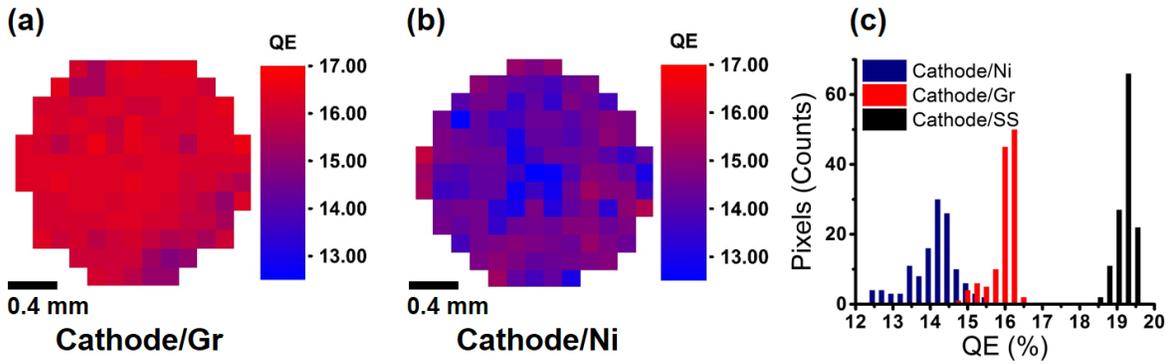

**Figure 2**. 405 nm quantum efficiency (QE) map of $K_2CsSb$ on (a) graphene and (b) Ni foil substrates in the sealed vacuum tube. The measurements were performed in reflection mode. Scale bars are 0.4 mm. (c) Statistics of QE in pixels of (a), (b), and $K_2CsSb$ on the chemically passivated stainless steel frame.

To address this critical point, we performed high spatial resolution QE mapping over smaller regions using a focused laser of 350 nm spot size. The photon energy was tunable thus we chose two typical values of 3.1 eV (400 nm) and 2.5 eV (500 nm). **Figure 3** shows the resulting QE maps as normalized to the solid supported regions, where a clear contrast is observed between the freely suspended and the supported photocathodes. The photocathodes suspended on graphene exhibit approximately 60 % and 80 % QE relative to the solid region at 3.1 eV and 2.5 eV respectively, illustrating the enhancement of QE by a reflective substrate. This is conceptually demonstrated in Figure 3(c) by calculating the optical absorption of a $K_2CsSb$ film freely suspended on graphene, compared to the case of a nearly ideal conductor (Aluminum) as a supporting substrate that yields notably enhanced absorption.

Given the 40 % geometrical mesh opening area, the results of e.g. Figure (2) can be explicitly decoupled by region, yielding QE ~ 14 % for the freely suspended photocathodes and ~ 17.5 % on the solid supported regions for 2.5 eV photons. The difference between the supported mesh region (17.5 %) and the SS reference region (19.3 %) is not surprising due to both the graphene film present on the mesh and the different surface treatments of these components (e.g. the very delicate SS mesh grids precluded chemical passivation) affecting the optical enhancement. More importantly however, the QE of the freely suspended photocathodes is consistent with that obtained on well-established transparent substrates when illuminated from the vacuum side (Supporting Information), thus strongly suggesting good compatibility between the photocathode and graphene. This is a significant advancement toward the realization of a photocathode interfacing with graphene for an atomically thin protection coating, which is our ultimate goal.

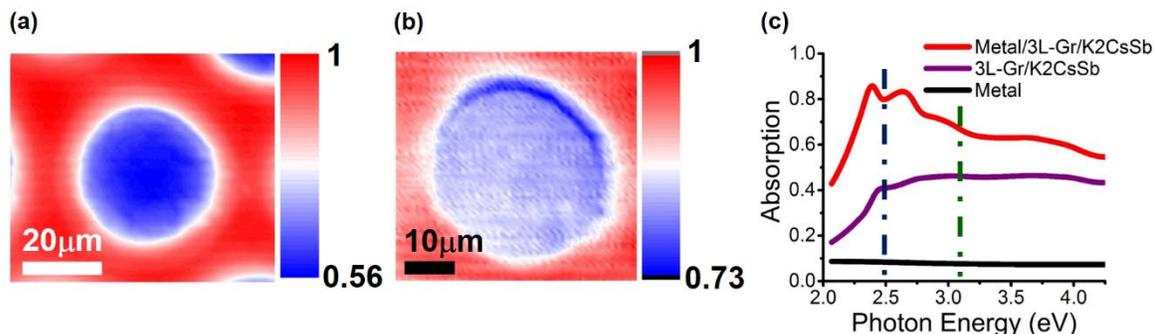

**Figure 3.** High resolution QE maps (a,b) of $K_2CsSb$ on graphene substrates deposited and sealed in vacuum. QE is normalized to the highest values on the supported regions as indicated in red. The measurements were performed in reflection mode. Excitation photons were (a) 2.5 eV (500 nm) at 1.7 µW, and (b) 3.1 eV (400 nm) at 170 nW. (c) Optical absorption calculated for the noted combinations of $K_2CsSb$, 3 monolayer graphene, and reflective aluminum ("Metal"). Dark blue and green dashed lines indicate the photon energy used in measurements of (a,b), respectively.

*2.1.3 Spectral response of quantum efficiency*

Distinct features in the overall shape of spectral response curves that pertain directly to the material's work function and electronic band structure can be used as an important tool to identify or compare photocathode materials. **Figure 4** shows the spectral responses for $K_2CsSb$ deposited on the SS frame, graphene-covered mesh areas, and Ni foil for photon energies ranging from 2-4.25 eV, as normalized to the SS given its established behavior. The spectral response for the graphene-covered mesh regions resembled that of the reference standard including the signature rising feature around 2.4 eV (black arrow). The inset is the enlarged low-energy cutoffs, which corresponds to the work function of photocathodes. Specifically, they were 1.84, 1.86, and 1.88 eV for photocathodes on graphene, SS, and Ni regions, respectively. They are all consistent with our previous study of 1.80-1.86 eV and reported values[35]. Note that the vertical axis of the inset is changed to square root of normalized QE for proper determination of the cutoff energy. It is also noted that the saturation of QE starting around 3.5 eV as seen in Figure 4 is a general characteristic shared among the family of the alkali antimonide photocathodes, where the primary cause is an onset of electron-electron (e-e) collisions. Electron-electron scattering events are prevented until the final states of both electrons can reside in the conduction band of $K_2CsSb$, thus for photon energies below approximately twice the band-gap energy the electrons excited from the top of the valence band, for example, do not have enough excess energy to scatter with another electron[37]. This leaves phonon scattering as the only scattering mechanism, so the QE continues to rise with increasing incident photon energy until the final states of two collisional electrons can both lie in the conduction band. This region corresponds to ~3.5 eV for $K_2CsSb$, and QE saturates or even drops as the incident photon energy increases[7]. Another contributing factor to the peak QE in the region of 3.0-3.5 eV is that optical absorption is highest in this region, as is optical penetration depth. The resulting shape and fine structure of the spectral response curve is therefore unique to the particular film in question (e.g., with a particular stoichiometry, crystallinity, thickness). The match of all of these features in the spectral response strongly suggests that $K_2CsSb$ films were successfully grown on the graphene regions without impact to the electronic structure. This is not necessarily trivial as certain substrates (e.g. copper (Cu)) are known to react with $K_2CsSb$, resulting in an order of magnitude lower QE compared to those deposited on SS[38, 39].

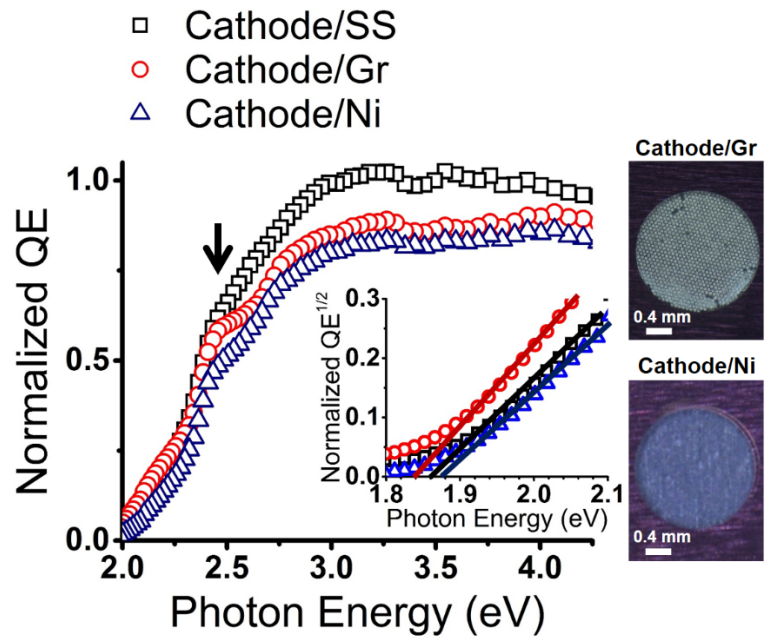

**Figure 4.** Normalized QE spectral response of $K_2CsSb$ photocathodes deposited on SS (black open squares), graphene-covered mesh (red open circles), and Ni foil (blue open triangle) in the sealed vacuum tube. QE is normalized to the photocathode on SS substrate. Black arrow indicates the position of signature rising feature. Inset #1 plot is the enlarged low-energy cutoff region of the spectra. Solid line is extrapolated to the horizontal axis to determine the cutoff energy. Inset #2 shows optical microscope images on the deposition side for $K_2CsSb$ on graphene (top) and Ni foil (bottom) substrates. Scale bars are 0.4 mm.

### 2.2 Material characterization of photocathodes on graphene substrates

While the QE and spectral response results using the sealed vacuum environment are sufficient to demonstrate the successful formation of $K_2CsSb$ on graphene, further characterization was performed during in-situ photocathode depositions using the Cornell High Energy Synchrotron Source (CHESS). An X-ray diffraction (XRD) result comparing bialkali antimonide photocathodes from depositions on graphene and a silicon (Si) reference substrate is shown in **Figure 5**(a). The observed peaks are assigned to (222), (022), (002), and (111), corresponding to theory for $K_2CsSb$ and confirming successful growth on the graphene substrate. The (002) peak was absent for the photocathode deposited on the Si reference, which is common for highly crystalline $K_2CsSb$[40]. The peaks are also 0.04-0.08 Å broader in the case of the $K_2CsSb$ on graphene substrate, as detailed in Table 1 for the various (HKL) crystal orientations. Both of these observations suggest a higher degree of crystallinity for the photocathode on the Si reference compared to that on the graphene substrate. X-ray reflectometry (XRR) indicated that surface roughness is not the origin of this difference[41]. Specifically, surface roughness of Si reference and graphene substrate prior to photocathodes deposition (2.7 ± 0.2 Å and 2.8 ± 0.1 Å, respectively) as well as that of deposited photocathodes (18.8 ± 0.2 Å and 17.0 ± 0.1 Å, respectively) were comparable (Supporting Information). Polymer residue on the surface of graphene substrates from the transfer process may be responsible for the lower degree of deposited photocathode crystallinity as more details are discussed in the last section. XRR also indicated that $K_2CsSb$ film could be thinner on graphene substrate using the same duration of growths. Wettability difference between deposited elements and Si reference or graphene substrates may be responsible for the observation. Optical transmission measurements of $K_2CsSb$ deposited on sapphire reference and graphene substrates in the sealed vacuum tube were consistent in that $K_2CsSb$ on graphene was thinner, however at a much lower degree[35]. Quantitative analysis of the difference is currently under investigation.

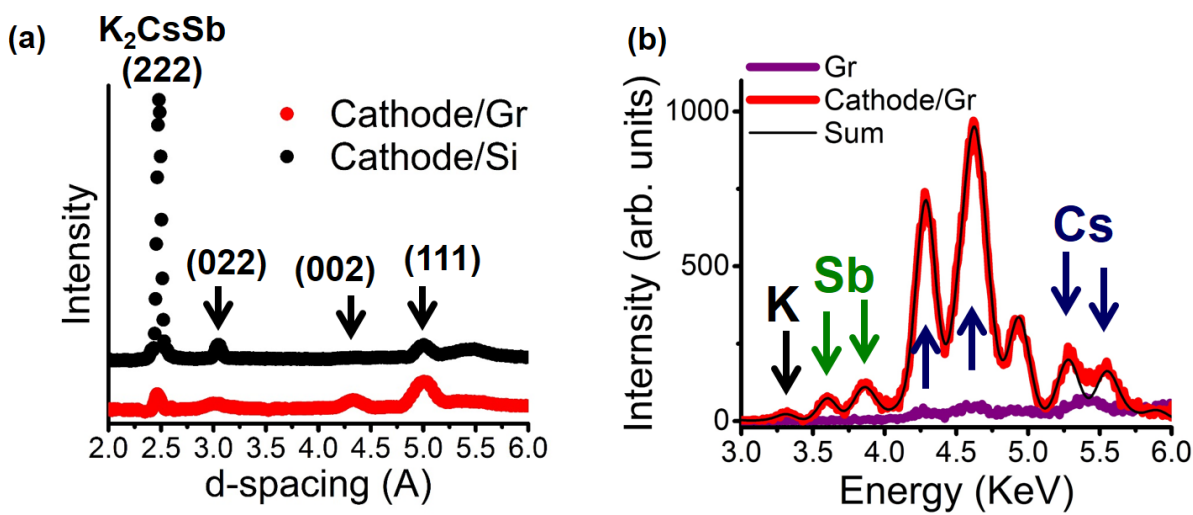

**Figure 5.** (a) X-ray diffraction (XRD) spectra of $K_2CsSb$ photocathodes deposited on Silicon (Si) (black squares) and graphene (red circles) substrates at the Cornell High Energy Synchrotron Source (CHESS). Intensity is normalized to the photocathode on Si substrate. (b) X-ray fluorescence (XRF) spectra of graphene substrate (purple) and $K_2CsSb$ photocathodes deposited on graphene substrates (red). Black, green, and blue arrows indicate the peak positions of potassium (K), antimony (Sb), and cesium (Cs), respectively. Black line is the sum of deconvoluted spectra used for quantitative analysis.

|  | d-spacing (Å) | Width (Å) | (HKL) | Theory (Å) |
|---|---|---|---|---|
| **Cathode/Si** | 2.48 | 0.05 | (222) | 2.49 |
|  | 3.05 | 0.08 | (022) | 3.05 |
|  | 5.00 | 0.16 | (111) | 4.98 |
| **Cathode/Gr** | 2.47 | 0.09 | (222) | 2.49 |
|  | 3.01 | 0.16 | (022) | 3.05 |
|  | 4.31 | - | (002) | 4.31 |
|  | 5.00 | 0.23 | (111) | 4.98 |

**Table 1.** List of d-spacing for observed peaks in comparison to the theoretical values of $K_2CsSb$. Corresponding crystal facet orientations and peak widths (full-width at half-maximum (FWHM)) are also shown.

We also performed X-ray fluorescence (XRF) spectroscopy to determine elemental composition of the photocathodes (Figure 4(b)). The photocathodes deposited on graphene clearly showed presence of potassium (K), antinomy (Sb), and cesium (Cs) as expected, where the unlabeled peak near 4.9 KeV corresponds to titanium (Ti) from the sample mount. Quantitative analysis by deconvolution of the spectra resulted in elemental composition of $K_{1.85}Cs_{1.08}Sb$. This is close to the ideal stoichiometry of $K_2CsSb$. The XRD and XRF results are consistent in that there are $K_2CsSb$ crystals present on graphene substrates, so they are likely responsible for the high QE we observed. The peak QE as well as its ratio between the photocathodes deposited here at CHESS on graphene and the reference is confirmed to be comparable to the ones deposited in the vacuum tube (Supporting Information).

### 2.3 Effect of graphene quality and thickness on quantum efficiency

*2.3.1 Relationship between quantum efficiency and graphene quality*

As a complementary investigation, $K_2CsSb$ photocathode was deposited and vacuum sealed using multiple graphene substrates with differences in annealing temperatures (350 °C and 500 °C), number of stacked monolayers (3 and 5), and also fabrication method (directly grown multilayers). The intention was to correlate any possible effect of residual water and/or graphene substrate thickness to the spectral response. Overall, all photocathodes exhibited similarly high QE of 14-17 % at ~3.1 eV on the graphene (mesh) regions, indicating that the above mentioned varied parameters had rather minimal effects on the QE. After careful characterization of graphene substrates using Raman spectroscopy, an interesting and insightful correlation between crystal quality of graphene substrates and QE of deposited photocathodes was found as follows.

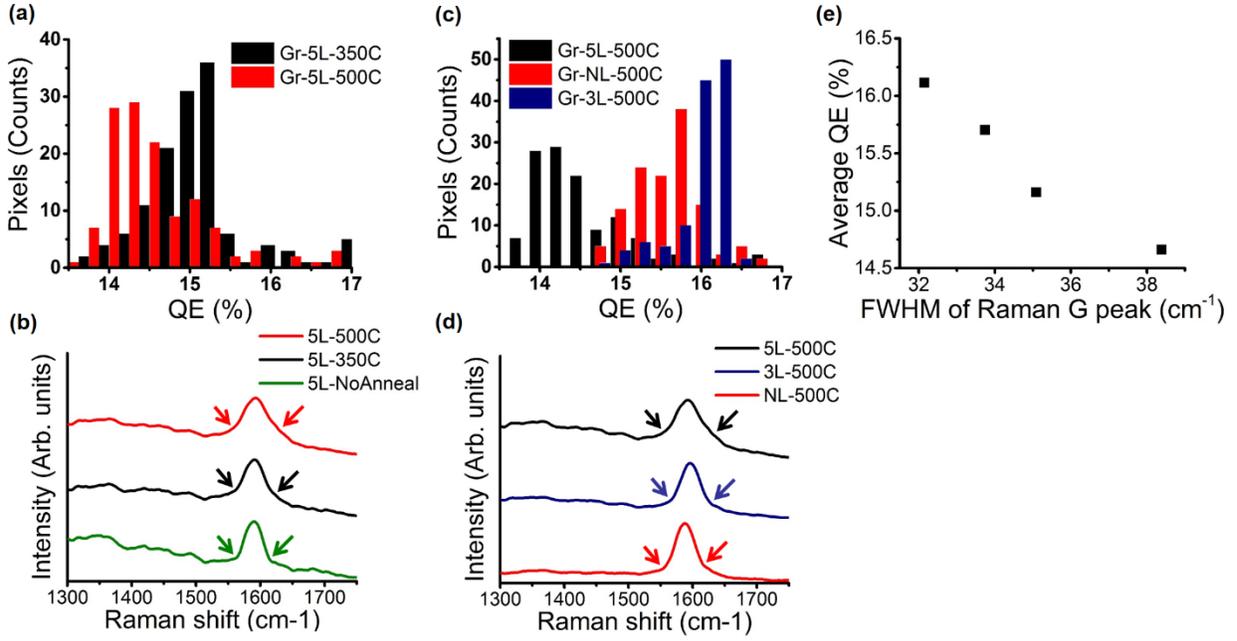

**Figure 6.** QE statistics of $K_2CsSb$ photocathodes on (a) 5L graphene substrates annealed at 350°C (black) and 500°C (red), (c) 5L (black), 3L (blue), NL (red) graphene substrates annealed at 500°C. (b), (d) Raman spectra of graphene substrates prior to $K_2CsSb$ deposition for the photocathodes in (a), (c), respectively. Arrows are eye guide for tails of G peaks at ~1590 cm$^{-1}$. (e) Average QE versus FWHM of Raman G peak extracted from (a)-(d).

**Figure 6** (a) shows the QE statistics of 0.15 mm x 0.15 mm pixels in the ~ 2 mm diameter $K_2CsSb$ photocathode deposited on 5 layer (5L) graphene that were annealed at 350 °C and 500 °C. Results indicated a slightly *higher* QE for $K_2CsSb$ deposited on 350 °C annealed graphene compared to the 500 °C sample. Raman measurements performed on the substrates prior to $K_2CsSb$ deposition indicated crystal structure defects as a possible origin for the observed results (Figure 6 (b)). Specifically, the G peak at ~1590 cm$^{-1}$ broadened as the annealing temperature increased. The FWHM increased from 29.96 cm$^{-1}$ for graphene without annealing to 35.09 cm$^{-1}$ and 38.39 cm$^{-1}$ for those annealed at 350 °C and 500 °C, respectively. Broadening of Raman peaks is indicative of an increase of crystal structural defects. The observed broadening of the G peak is consistent with Suzuki *et al.*[42], which reports that interaction between residual water at the interlayer and graphene after the wet-transfer process induces atomic structural defects at annealing temperatures higher than 300 °C. Their study is based on thorough electrical transport, Raman spectroscopy and X-ray photoelectron spectroscopy results. It is worthwhile mentioning that a D peak at ~1350 cm$^{-1}$ was not observed even after the noted broadening of the G peak for graphene annealed at 500 °C. This result suggests that the amount of crystal structural defects to have an impact on the QE of the deposited K$_2$CsSb could be quite small. Furthermore, it also suggests that the in-situ pre-cleaning temperature of 350 °C is high enough, thus increasing the annealing temperature can have a negative impact on the QE of deposited photocathodes.

The conclusion drawn from investigating the graphene thickness effect on the QE of deposited K$_2$CsSb was consistent with that of varying the graphene annealing temperature. That is, eliminating the residual water at the interlayer is an important factor for improving the substrate quality and resultant QE. Figure 6 (c) shows QE statistics for pixels of K$_2$CsSb deposited on graphene substrates with three and five layers (3L and 5L). All of the graphene substrates were annealed at 500 °C. The higher QE observed for the 3L case can be interpreted by the relatively less residual water-induced defects upon annealing as compared to that of the 5L, as is evident from the G peak FWHM broadening in the Raman spectra. It broadened from 29.96 cm$^{-1}$ for graphene without annealing to 32.15 cm$^{-1}$ and 38.39 cm$^{-1}$ for 3L and 5L, respectively as shown in Figure 6 (d). Directly grown multilayer (NL, thickness of ~2 nm thus ~3 layers) undergoes only one wet-transfer process thus a minimal exposure to the water is the most probable cause for it to yield the similar QE as that of 3L. The FWHM of the Raman G peak was 33.74 cm$^{-1}$, which is also comparable to that of 3L. Figure 6 (e) effectively summarizes the trend of higher QE achieved from K$_2$CsSb deposited on graphene substrates with narrower FWHM of G peaks in the Raman spectra.

*2.3.2 Photoemission through graphene and effect of thickness*

All of the photoemission measurements thus far were performed from the photocathode sides. However we also attempted to measure photoemission through the graphene side as it directly relates to the ultimate application goals, i.e. photoemission through graphene is required to encapsulate a photocathode to enhance its lifetime. Unfortunately no detectable photoemission from the graphene side was observed. We used 238 nm (5.2 eV) ultraviolet (UV) LED excitation in addition to the lamp for this purpose based on the expected increase of work function when photocathode is interfaced with graphene[35]. We also applied increased anode voltage to compensate the electric field shielding due to mesh grid geometry (Supporting Information). We speculate that residue of polymer used for graphene transfer process is primary responsible for this as follows.

First, we have previously experimentally demonstrated that when monolayer graphene forms a clean interface with metallic photocathodes such as Cu, the encapsulation does not decrease QE and indeed increase its robustness to the environment[34]. Moreover, we have theoretically shown that when trilayer graphene forms clean interface with photocathodes, the peak QE is well within the detectable range (several %). Therefore the present results suggest that either our trilayer graphene or the interface are not atomically clean. A possible cause for this is polymer residue from the graphene transfer process. Specifically, reported literature as well as our low energy electron diffraction (LEED) investigation combined with AFM suggest that there is <1nm PMMA residue on the surface of graphene even after its removal in acetone bath[43]. It is worthwhile noting that this is not identifiable in a conventional optical transmission measurements. This residue appears persistent regardless of the number of rinse baths (up to 5 times) or temperature increases during rinsing (up to 80 °C). However, we recently confirmed by LEED that the residue could be removed by annealing it above 350 °C in vacuum or in presence of hydrogen gas. In this study, graphene substrates were pre-cleaned for several hours above 350 °C in vacuum prior to the photocathode deposition for the vacuum tube or the material characterization. This should have removed a majority of PMMA residue on the top graphene surface, which comes in direct contact with the deposited photocathodes. Interlayers of stacked graphene, on the other hand, most likely contained PMMA residue and we speculate this increased the thickness of our trilayer graphene substrates. Quantum tunneling probability decays exponentially with increased distance and if there were 2 nm worth of PMMA residue at the interlayer of our graphene substrates, it will be enough to hinder photoemission through graphene. A separate study on graphene prepared by thorough PMMA removal (PMMA residue removed at interlayers also) is currently underway to address this issue.

## 3. Conclusions

In summary, bialkali antimonide photocathodes were successfully grown on freely suspended graphene substrates of few-monolayer thickness. The large-area uniformity, quantum efficiencies, specific spectral response features, and comparisons to experimental reference standards and calculation results all indicate a high compatibility of the photocathodes with thin graphene layers. Additional material characterization using XRD and XRF directly revealed grains of $K_2CsSb$ in such photocathodes when deposited on graphene, further supporting this compatibility.  Variations in the number of graphene layers, their stacking methods, and additional pre-deposition annealing variations had small effects on the QE magnitudes of the resultant photocathodes, yet highlighted areas for measurable improvements by minimizing defects in the graphene films.  The results demonstrate that high QEs can be achieved for $K_2CsSb$ on thin graphene substrates that are as thin as three atomic layers. This is a promising step toward fabrication of photocathodes with enhanced lifetimes *via* atomically thin protection layers as well as photocathodes on optically transparent yet electrically conductive substrates.

## Experimental Section

*Synthesis and transfer of graphene:* Graphene was synthesized *via* chemical vapor deposition (CVD) using methane gas as the carbon source and copper (Cu) foils as substrates. Monolayers were grown to stack three and five layer films, while multilayer deposited films were used directly. For wet-transfer of CVD graphene onto mesh grids, poly(methyl methacrylate) (PMMA) was used as a mechanical support and removed by acetone rinsing afterwards. Intermediate transfer onto rigid substrates for a stacking purpose was crucial for increasing the transfer yield. Some samples were annealed up to 500 ºC in ultrahigh vacuum (UHV) after the final transfer to remove residual moisture.

*Deposition of Bialkali antimonide photocathodes for vacuum tube:* Graphene films on TEM SS mesh grids (SPI Supplies / Structure Probe, Inc. S400, 34 μm holes, 17 μm thickness) and Ni reference foil (Alfa Aeser 12722, 99 %, annealed) were installed into SS foil frames (SS304) for bialkali antimonide photocathode deposition at Photonis USA. The SS frames had ~2 mm holes to allow placement and illumination of the samples. All vacuum envelope components including the SS frame with graphene substrates were pre-cleaned at 350 ºC in UHV prior to *in-situ* photocathode deposition. While monitoring the sensitivity of the photocathode films, the components K, Cs, and Sb were deposited on the SS foil, Ni reference, and graphene substrates *via* thermal evaporation to achieve typical stoichiometry of $K_2CsSb$ with thickness of ~25-30 nm. The vacuum-sealed package consisted of sapphire windows on both sides of the photocathode assembly with metal traces patterned on the windows to establish an extracting electric field.

*Photoemission measurement of bialkali antimonide photocathodes in vacuum tube:* A 405 nm light emitting diode (LED) (for whole area QE maps), xenon (Xe) lamp with monochromator (for spectral responses), and Fianium Whitelase tunable laser (400-2400 nm, repetition rate 40 MHz) equipped with a Fianium AOTF (Acousto-Optic Tunable Filter) system (for high spatial resolution QE maps) were used as light sources for photoemission measurements. The focused spot size was ~0.20 mm, ~1.5 mm, <350 nm for the 405 nm LED, lamp, and tunable laser, respectively. Anode traces on the sapphire windows were sufficiently biased with respect to the photocathode assembly to overcome space-charge effects and collect photoelectrons in all cases.

The quantum efficiency was calculated using the known power of incident light at each wavelength, as obtained from a calibrated reference diode. The energy of the incident light was scanned from 2 to 4.25 eV while the corresponding photocurrent was recorded to obtain spectral response of the photocathodes. A home-built confocal microscopy system with a scanning mirror that allows for precise location of the focal point onto the sample surface was used for high spatial resolution QE maps.

*Material characterization of bialkali antimonide photocathodes:* In-situ X-ray diffraction (XRD) growth studies on $K_2CsSb$ were performed at the Cornell High Energy Synchrotron Source (CHESS) beamline G3 using photon energy of 13 KeV ($\lambda = 0.95$ Å). The thin film growth was performed in a custom-built ultrahigh vacuum chamber with a base pressure of low $10^{-10}$ Torr. Graphene was grown by CVD and transferred onto Si substrates freshly etched by diluted hydrofluoric acid (HF) to the remove native oxide. The reference Si substrates were etched by HF and rinsed and stored in deionized water before being loaded into the chamber. Both substrates were loaded into the growth chamber and annealed at 550 °C for 1 hour. Co-evaporation of K, Cs and Sb using pure metallic sources was used to fabricate $K_2CsSb$ photocathodes. The evaporation rate was controlled by adjusting the current of the evaporators and was measured with a quartz crystal microbalance (QCM) placed alongside the sample. Alkali and antimony sources were turned on simultaneously, and the rates of the three were set to match the stoichiometry of $K_2CsSb$ based on real-time X-ray fluorescence (XRF) analysis. During deposition, the substrate temperature was set to about 90 °C. The X-ray reflectometry (XRR) and XRD data were measured using a 4 axis diffractometer with a Pilatus 100 K X-ray camera mounted 100 cm downstream from the substrate. XRR measurements were performed by scanning the 2θ angle from 0° to 6° and XRD measured with a 2θ range from 5° to 30°. The XRF spectra were measured by a vortex multi-cathode X-ray detector mounted 45° with respect to the sample surface normal and approximately 25 cm away from the sample. The photocurrent was monitored during the growth by a 532 nm diode laser and the signal was collected using a Keithley 6517B electrometer. After growth the spectral response of each sample from 250 nm to 700 nm was measured using an optical system consisting of a laser driven light source (LDLS) and a Cornerstone monochromator.

***Absorption calculation of bialkali antimonide photocathodes:*** NKD Gen software (available from University of Barcelona) that is based on conventional transfer matrix methodology was used for the calculation. Established literature values were used for optical constants (n,k) of $K_2CsSb$, graphene, and aluminum metal, and thickness of 20 nm and 1.07 nm were used for the $K_2CsSb$ and 3 layer (3L) graphene, respectively.

## Acknowledgments

Authors acknowledge financial supports from the Los Alamos National Laboratory (LANL) Laboratory Directed Research and Development (LDRD) Program through Directed Research (DR) "Applied Cathode Enhancement and Robustness Technologies (ACERT)" (Project #20150394DR). Studies were performed, in part, at the Center for Integrated Nanotechnologies, an Office of Science User Facility operated for the U.S. Department of Energy (DOE), Office of Science. LANL, an affirmative action equal opportunity employer, is operated by Los Alamos National Security, LLC, for the National Nuclear Security Administration of the U.S. Department of Energy under contract DE-AC52-06NA25396. The Cornell High Energy Synchrotron Source (CHESS) is supported by the National Science Foundation and the National Institutes of Health/National Institute of General Medical Sciences under NSF Award Nos. DMR-0936384 and DMR-1332208. H.Y. acknowledges Dr. Andrew M. Dattelbaum of LANL and Prof. Gautum Gupta of University of Louisville for their generous supports that enabled this work to be performed without an administrative delay. Work at Argonne National Laboratory was supported by the U.S. DOE, Office of Science, Office of Basic Energy Sciences and Office of High Energy Physics under contract DE-AC02-06CH11357.

## Conflict of Interest

N.A.Moody owns the patent US 8,823,259 on a concept of graphene protection of chemically reactive films.

# Supporting Information

## Free-standing bialkali photocathodes using atomically thin substrates

*Hisato Yamaguchi[#,\*], Fangze Liu[#], Jeffrey DeFazio, Mengjia Gaowei, Claudia W. Narvaez Villarrubia, Junqi Xie, John Sinsheimer, Derek Strom, Vitaly Pavlenko, Kevin L. Jensen, John Smedley, Aditya D. Mohite, Nathan A. Moody[\*]*

## Contents

*S1: Raman spectroscopy on monolayer chemical vapor deposition (CVD) graphene*

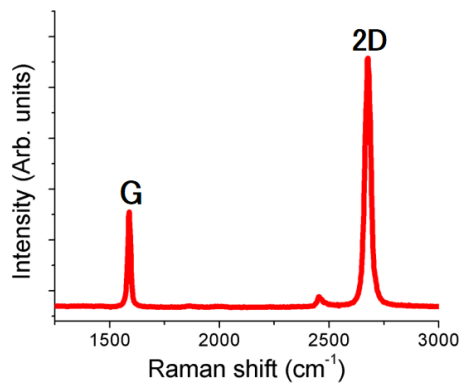

Figure S1 - Raman spectrum of typical monolayer CVD graphene after transferred onto $SiO_2$/Si substrate using wet-based method [S1].

*S2: Atomic force microscopy (AFM) on monolayer CVD graphene*

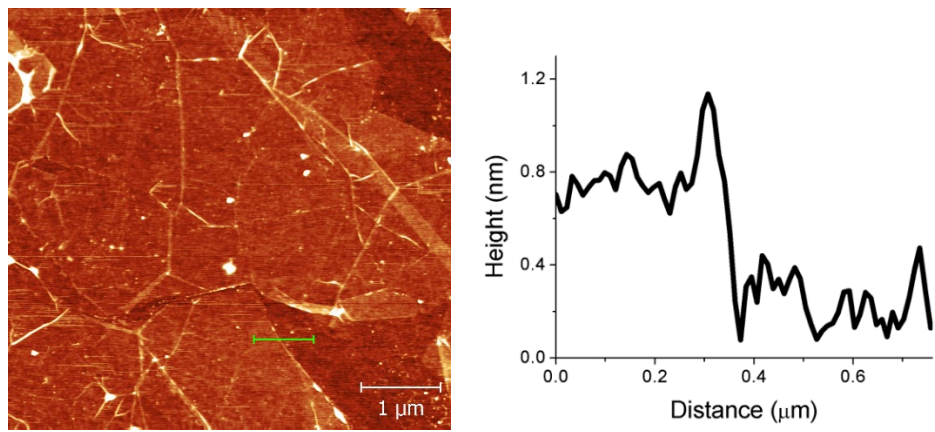

Figure S2 - AFM image (left) and extracted height profile (right) of typical monolayer CVD graphene at the edges after transferred onto $SiO_2$/Si substrate using wet-based method [S1]. Green line in AFM image indicates the position where height profile was taken. Scale bar is 1 μm.

*S3: Scanning electron and optical microscopy on monolayer CVD graphene*

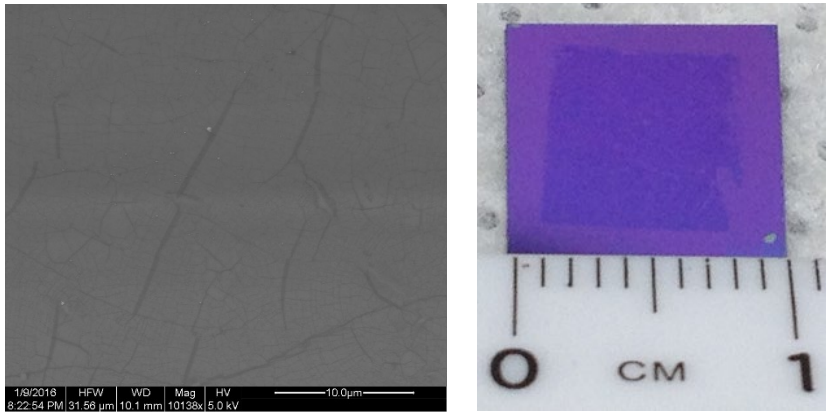

Figure S3 - Scanning electron (left) and optical microscopy (right) images of typical monolayer CVD graphene after transferred onto SiO$_2$/Si substrate using wet-based method. Scale bar is 10 μm for the scanning electron microscopy image.

*S4: QE spectral response of photocathode on a thick transparent substrate*

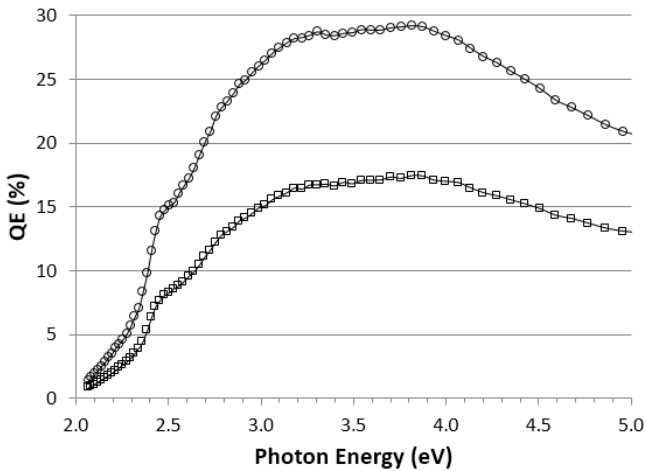

Figure S4 - QE spectral response of K$_2$CsSb photocathode on a thick transparent substrate (sapphire) when illuminated from the substrate side (circles) or the vacuum side (squares).

*S5: QE spectral response of photocathodes used for material characterization*

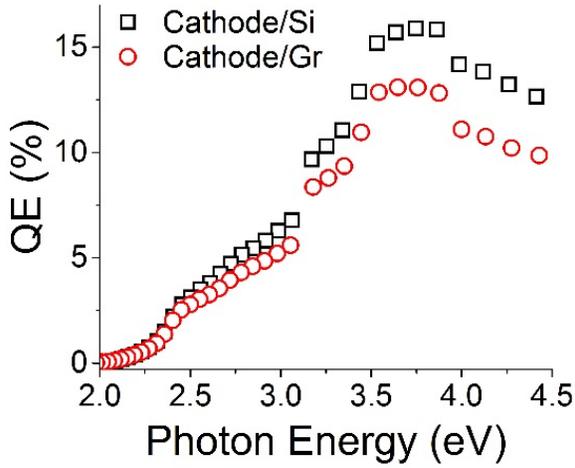

Figure S5 - QE spectral response of $K_2CsSb$ photocathodes deposited on silicon (Si) and graphene substrates for material characterization purpose at the Cornell High Energy Synchrotron Source (CHESS).

*S6: X-ray reflectometry (XRR) spectra of photocathodes and substrates*

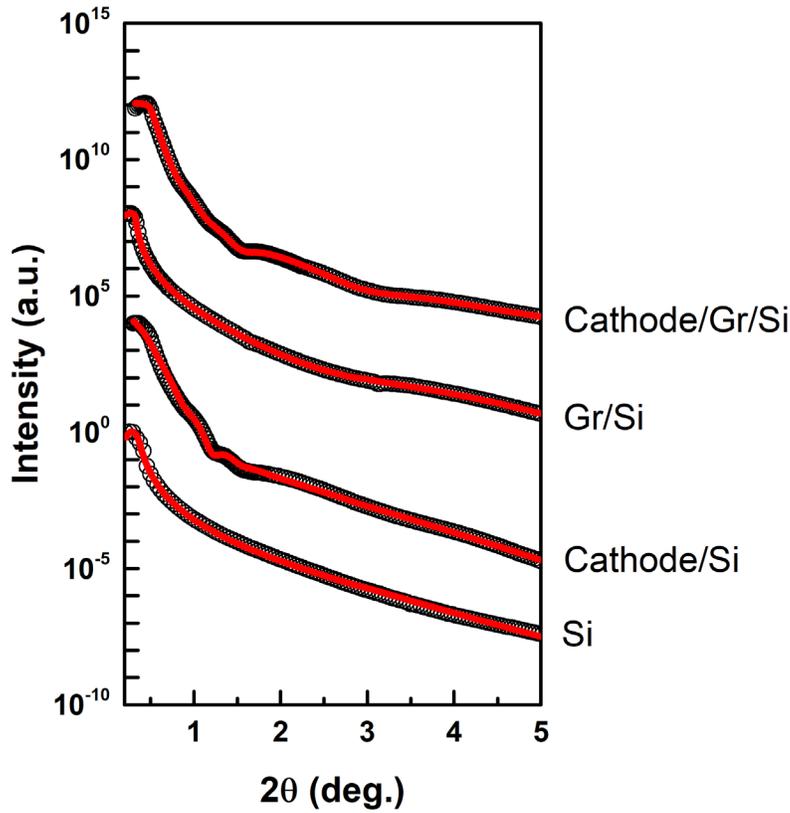

Figure S6 - XRR spectra of $K_2CsSb$ photocathodes deposited on silicon (Si) and graphene substrates (Gr/Si) for material characterization purpose at the CHESS. Points are XRR experiment data, and solid lines are fitted results using Parratt's recursion [S2]. The plots are offset for clarity.

*S7: Electric field screening by a mesh grid geometry*

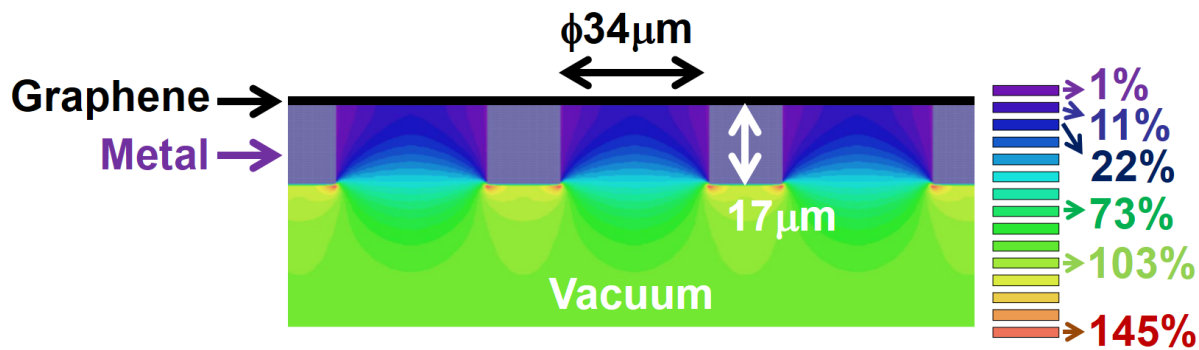

Figure S7 - Calculated electric field distribution around a mesh grid. Mesh grid geometry is consistent with the experiment; 34 μm diameter holes and 17 μm thickness. Relative electric field at the graphene surface to that of flat surface is 15 % at the center of holes (dark blue) and ~1 % at the edges (purple). HiPhi from Field Precision is used for the calculation [S3].